\RequirePackage{lineno}

\documentclass[onecolumn,superscriptaddress,showkeys,bibnotes]{revtex4}

\usepackage{lineno}

\usepackage{amsmath}
\usepackage{subfig}
\usepackage{graphicx}

\begin{document}

\title{A Parameterized Energy Correction Method for Electromagnetic Showers in BGO-ECAL of DAMPE}

\author{Chuan Yue \footnote{The corresponding author (yuechuan@pmo.ac.cn).}}
\affiliation{Key Laboratory of Dark Matter and Space Astronomy, Purple Mountain Observatory, Chinese Academy of Sciences, Nanjing 210008, China}
\affiliation{University of Chinese Academy of Sciences, Yuquan Road 19, Beijing 100049, China}

\author{Jingjing Zang  \footnote{The corresponding author (zangjj@pmo.ac.cn).}}
\affiliation{Key Laboratory of Dark Matter and Space Astronomy, Purple Mountain Observatory, Chinese Academy of Sciences, Nanjing 210008, China}

\author{Tiekuang Dong}
\affiliation{Key Laboratory of Dark Matter and Space Astronomy, Purple Mountain Observatory, Chinese Academy of Sciences, Nanjing 210008, China}

\author{Xiang Li}
\affiliation{Key Laboratory of Dark Matter and Space Astronomy, Purple Mountain Observatory, Chinese Academy of Sciences, Nanjing 210008, China}

\author{Zhiyong Zhang}
\affiliation{State Key Laboratory of Particle Detection and Electronics, University of Science and Technology of China, Hefei 230026, China}

\author{Stephan Zimmer}
\affiliation{Department of Nuclear and Particle Physics, University of Geneva, CH-1211, Switzerland}

\author{Wei Jiang}
\affiliation{Key Laboratory of Dark Matter and Space Astronomy, Purple Mountain Observatory, Chinese Academy of Sciences, Nanjing 210008, China}
\affiliation{School of Astronomy and Space Science, University of Science and Technology of China, Hefei 230026, China}

\author{Yunlong Zhang}
\affiliation{State Key Laboratory of Particle Detection and Electronics, University of Science and Technology of China, Hefei 230026, China}

\author{Daming Wei}
\affiliation{Key Laboratory of Dark Matter and Space Astronomy, Purple Mountain Observatory, Chinese Academy of Sciences, Nanjing 210008, China}
\affiliation{School of Astronomy and Space Science, University of Science and Technology of China, Hefei 230026, China}

\begin{abstract}
DAMPE is a space-based mission designed as a high energy particle detector measuring cosmic-rays and $\gamma-$rays which was successfully launched on Dec.17, 2015. The BGO electromagnetic calorimeter is one of the key sub-detectors of DAMPE for energy measurement of electromagnetic showers produced by $e^{\pm}/{\gamma}$. Due to energy loss in dead material and energy leakage outside the calorimeter, the deposited energy in BGO underestimates the primary energy of incident $e^{\pm}/{\gamma}$. In this paper, based on detailed MC simulations, a parameterized energy correction method using the lateral and longitudinal information of electromagnetic showers has been studied and verified with data of electron beam test at CERN. The measurements of energy linearity and resolution are significant improved by applying this correction method for electromagnetic showers.
\end{abstract}


\maketitle

\section{Introduction}
\begin{figure}[!ht]
  \centering  
  \includegraphics[width=0.8\textwidth]{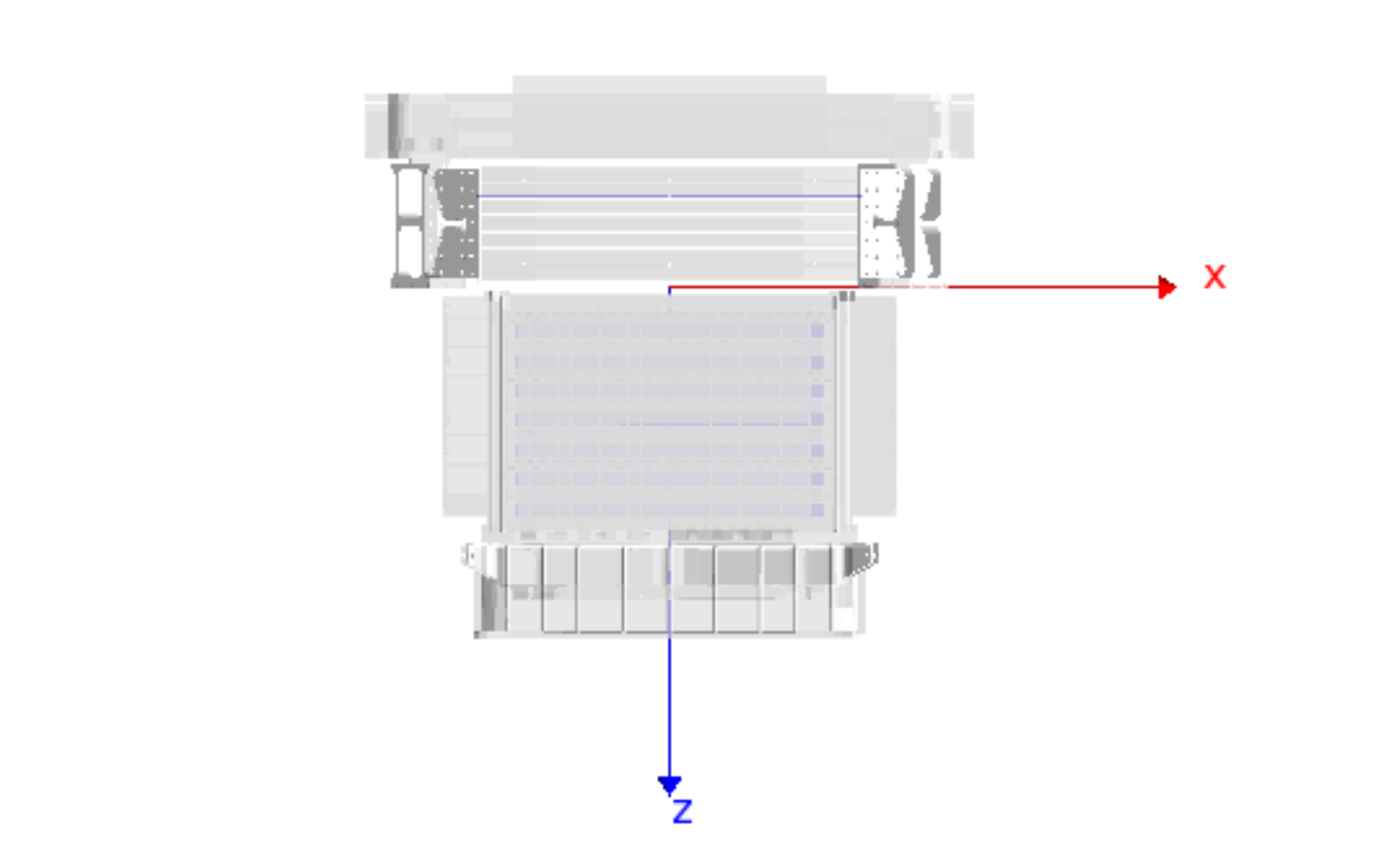}
  \caption{The scientific payload of DAMPE, which consists of four sub-detectors including PSD, STK, BGO-ECAL and NUD from up to down.}
  \label{fig-payload}
\end{figure}

The DArk Matter Particle Explorer (DAMPE), an orbital mission supported by the strategic priority science and technology projects in space science of the Chinese Academy of Science \cite{ChangJ2014}, has been successfully launched into a sun-synchronous orbit at the altitude of 500 km on Dec.17 2015. By measuring cosmic rays and $\gamma-$rays in a wide energy range, the DAMPE offers a new opportunity for advancing our knowledge of dark matter, propagation mechanisms of cosmic rays and high energy astronomy. The scientific payload of DAMPE consists of four sub-detectors\cite{DmpMission} including a Plastic Scintillator strip Detector (PSD), a Silicon-Tungsten tracKer-converter (STK), a Bismuth-Germanium Oxide (BGO) imaging Energy CALorimeter (ECAL), and a NeUtron Detector (NUD) (see  Fig.\ref{fig-payload}). All sub-detectors work together to measure particle-id, track and energy information of the incident cosmic-ray particles. 

The BGO-ECAL is the main sub-detector for energy measurement, covering a large energy range from 5 GeV to 10 TeV with an energy resolution better than $1.5\%$ at 800 GeV \cite{ChangJ2009}. The BGO-ECAL is designed as a total-absorption electromagnetic calorimeter of about 31 radiation-length, composed of 14 layers, each layer consists of 22 BGO crystals ($25 \times 25 \times 600 mm^{3}$) placed in a hodoscopic configuration \cite{ZhangYL2012}. Apart from measuring the energy depositions due to the electromagnetic showers produced by $e^{\pm}/{\gamma}$ , the ECAL images their shower development profiles, thereby serving as an important hadron/lepton discriminator. 

The overall raw energy is taken to be the sum of the deposited energy in each crystal of the calorimeter. However, the total deposited energy in the BGO-ECAL cannot be straightforwardly taken as the primary energy of incident particle, due to non-negligible energy losses because of dead material and leakage outside the calorimeter. In previous work, it has been studied that the energy loss in the gaps between BGO crystals is sensitive to the impact of the particle with respect to the boundaries of each cell\cite{LiZY2016}. In this paper, a parameterized correction method based on the electromagnetic shower developments in ECAL is studied to estimate the primary energies of incident particles.

\section{DAMPE Simulation}
The DAMPE simulation is based upon the GEANT4 toolkit \cite{Allison2006}, a Monte-Carlo (MC) simulation software widely used in high energy physics experiments, to handle particle generation, propagation and interaction through the instrument with certain physical models. The whole simulation package is implemented in a GAUDI-like software framework \cite{Tykhonov2015,WangC2016}, which would produce collections of energy hits for each sensitive detector. A digitization algorithm has been developed and used for converting energy hits into Analog-to-Digital Converter (ADC) counts which have same format as the raw flight data. To reflect the electronic response of each sub-detector, the digitization algorithm takes into account the on-orbit calibration results such as pedestal noise, PhotoMultiplier Tube (PMT) gains, non-uniformities etc. With the on-orbit calibration of trigger threshold, real data trigger logic is then applied to the MC data. In this way, the MC data can be processed by the same reconstruction algorithms as the real data.

\begin{figure}[!ht]
  \centering  
  \includegraphics[width=.8\textwidth]{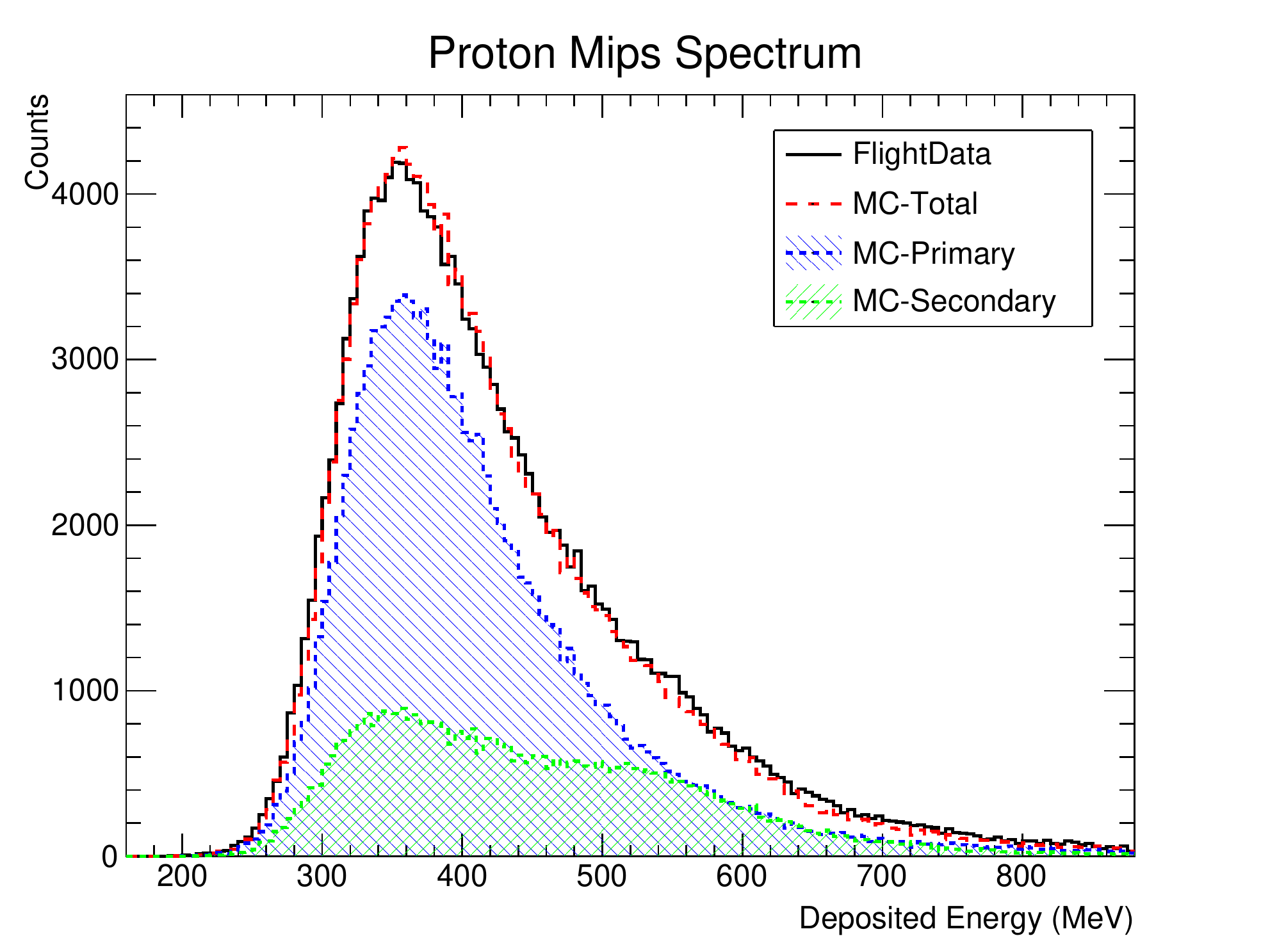}
  \caption{The spectrum of proton MIPs events measured by DAMPE. It is evident that the on-orbit flight-data (black) match well with the simu-data(red). The spectra of primary protons and secondary protons used in simulations are shown for reference.}
  \label{fig-mipspectrum}
\end{figure}

Before the MC simulation can be used for analysis, the physical processes and the geometry implementation have to be validated by comparing the simulation results with flight data. On orbit, the cosmic-ray proton Minimum Ionizing Particle (MIP) events at low latitudes (${\pm}20^{\circ}N$, discarding the South Atlantic Anomaly (SAA) region) has been selected to calibrate the ADC-Energy response for each BGO crystal. By comparing the proton MIPs spectrum of flight-data with simu-data, the BGO-ECAL simulation package can be proved to be reliable. A cosmic-ray proton flux model at low latitudes, including primary protons and secondary ones measured by AMS-01 experiment \cite{AMS01} , has been used as the input spectrum in the MIPs simulation. In Fig.\ref{fig-mipspectrum}, the comparisons of the proton MIPs spectrum show good agreement between flight-data and simu-data. More details about the validation of the DAMPE simulation will be published elsewhere. 

In this paper, several isotropic electron sources with kinetic energies from 1 GeV to 5 TeV have been simulated and digitized by the DAMPE simulation package. The simulation data were then reconstructed and selected by the same algorithms used in flight-data analysis. The selected electron datasets were used for studying the energy response of electromagnetic showers in BGO-ECAL.

\section{Energy Correction Method}
As mentioned above, for $e^{\pm}/{\gamma}$, the electromagnetic shower is well contained by the 31 radiation-length calorimeter in most cases and most of incident energy would be deposited in the calorimeter. The energy underestimation is mainly due to energy loss in the dead material of BGO-ECAL, such as the carbon-fibers and rubbers used as the calorimeter support structure. Also, the deposited energy in materials in front of the calorimeter (PSD, STK and the tungsten that used to convert the gamma-rays) cannot be neglected. For the incident $e^{\pm}/{\gamma}$ with energy higher than hundreds of GeV, the rear energy leakage must be taken into account as well.

\subsection{Lateral Correction}
\begin{figure*}[!ht]
  \centering
  \subfloat[10 GeV MC electron]{
    \includegraphics[width=.32\textwidth]{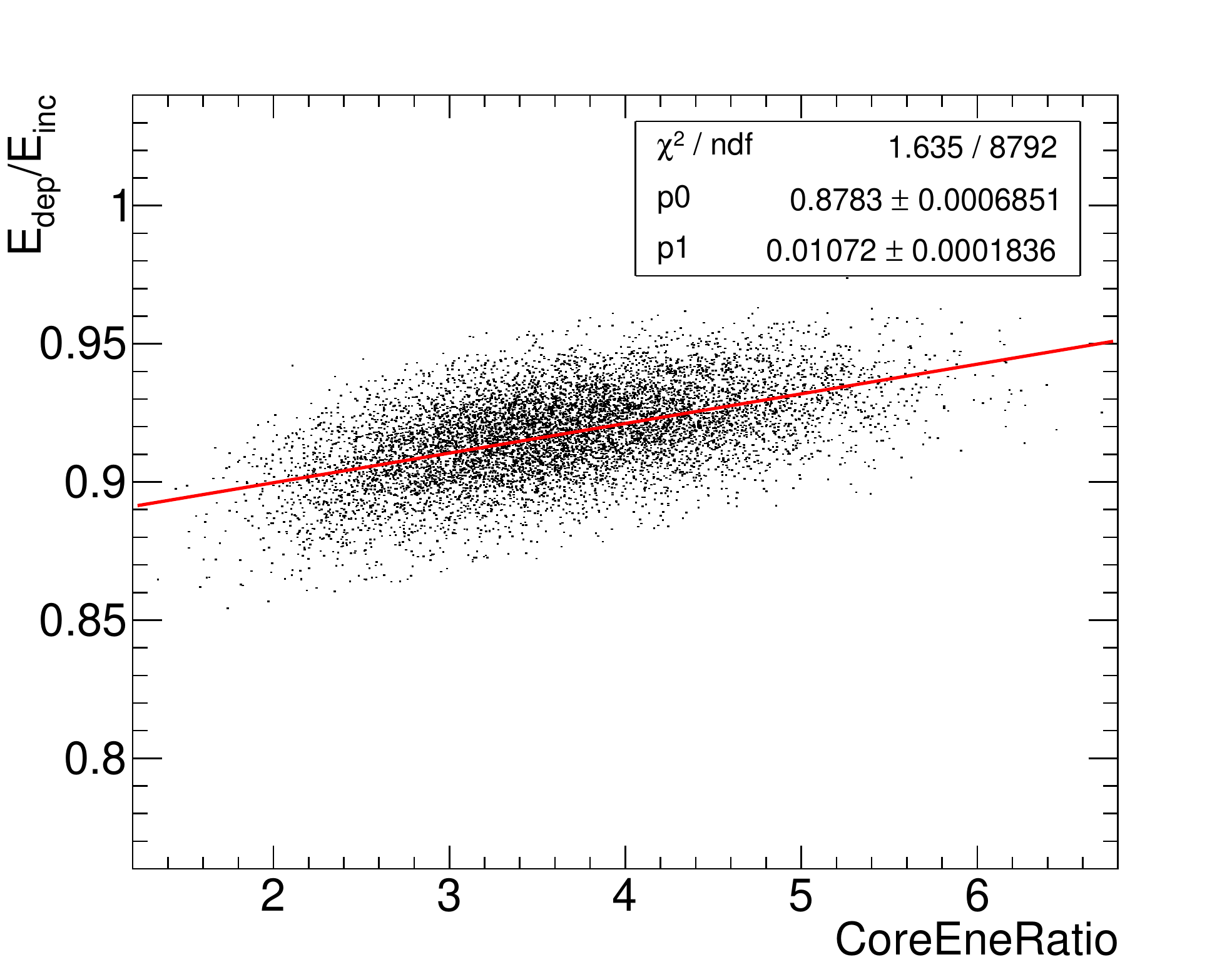}
  }
  \subfloat[100 GeV MC electron]{
    \includegraphics[width=.32\textwidth]{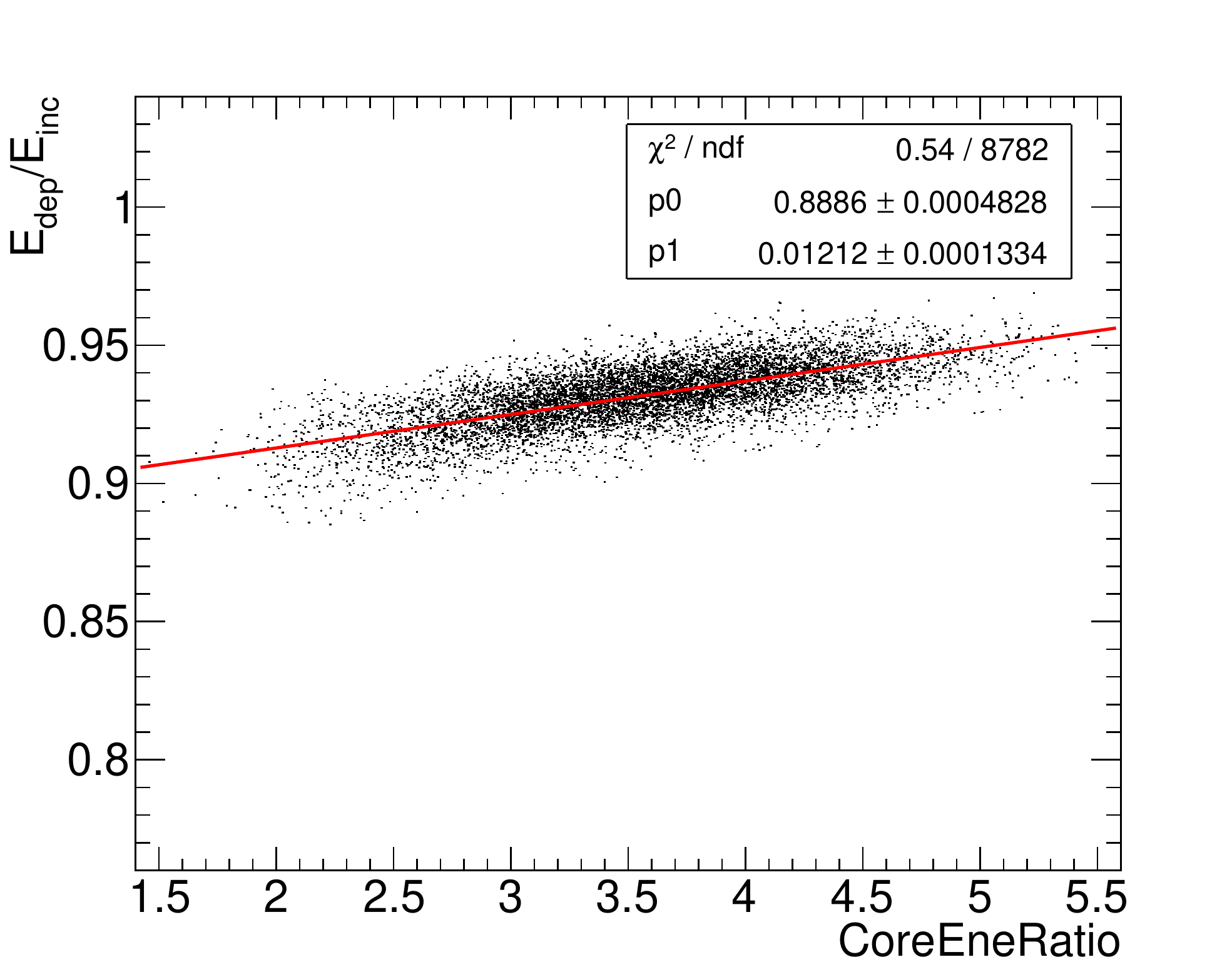}
  }
  \subfloat[1000 GeV MC electron]{
    \includegraphics[width=.32\textwidth]{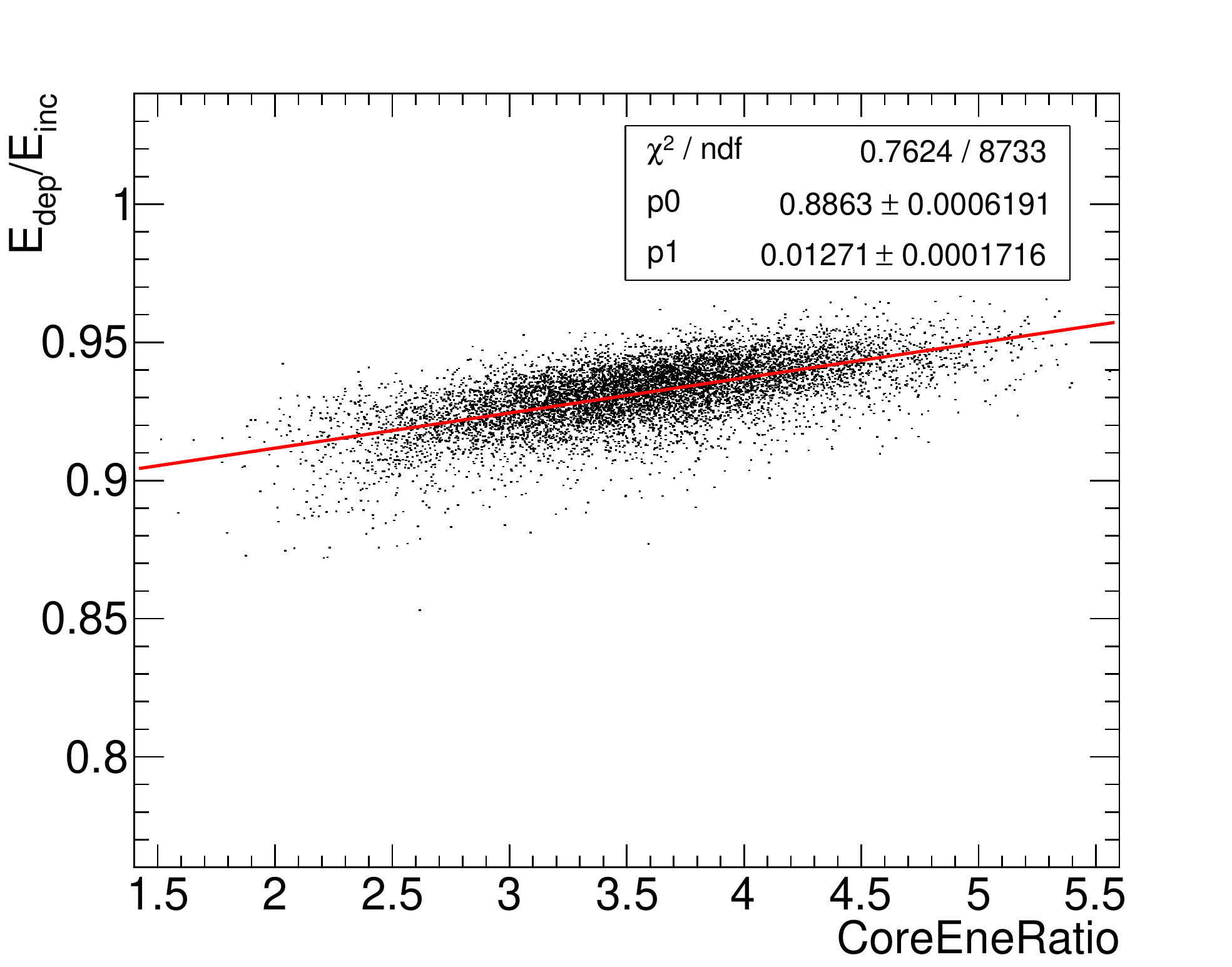}
  }
  \caption{The variations of $EdepRatio$ along with $CoreEneRatio$ for (a)10GeV, (b)100GeV and 1000GeV MC electrons. The linearity relation between $EdepRatio$ and $CoreEneRatio$ are fitted as red line.}
  \label{fig-coreratio}
\end{figure*}

To correct for energy loss in the gaps between BGO crystals, a method called S1/S3 from \cite{S1S3} has been frequently used in some other similar experiments \cite{Adloff2013}. This method takes into account the dead material in small gaps between neighboring crystals. Given the large size of single BGO crystal used in DAMPE, we constructed a correction variable called $CoreEneRatio$ adapted from the S1/S3 method: the ratio of the sum of the maximum energy in each layer to the residual deposited energy multiplied by a angle factor, defined as the following: 
\begin{linenomath}
\begin{equation}
\begin{split}
& CoreEneRatio =
\\
& \frac{\sum{MaxBarE_{i}}}{\sum{LayerE_{i}}-\sum{MaxBarE_{i}}} \cdot \frac{1}{cos{\theta}}
\end{split}
\label{eq-coreratio}
\end{equation}
\end{linenomath}
where $MaxBarE_{i}$ is the maximum energy of layer $i$, $LayerE_{i}$ is the total energy of layer $i$ and $\theta$ is the measured angle with z-axis. Fig.\ref{fig-coreratio} shows the dependance of the energy deposition ratio (the ratio of the deposited energy to the primary energy) with  $CoreEneRatio$ for 10GeV, 100GeV and 1000GeV MC electrons. It suggests that smaller $CoreEneRatio$ indicates more energy loss in the gaps between BGO crystals. This good linear relationship can be used to correct the energy loss in the dead material which is related to the particle injecting position.

\subsection{Longitudinal Correction}
\begin{figure*}[!ht]
  \centering
  \subfloat[5 GeV MC electron]{
    \includegraphics[width=.40\textwidth]{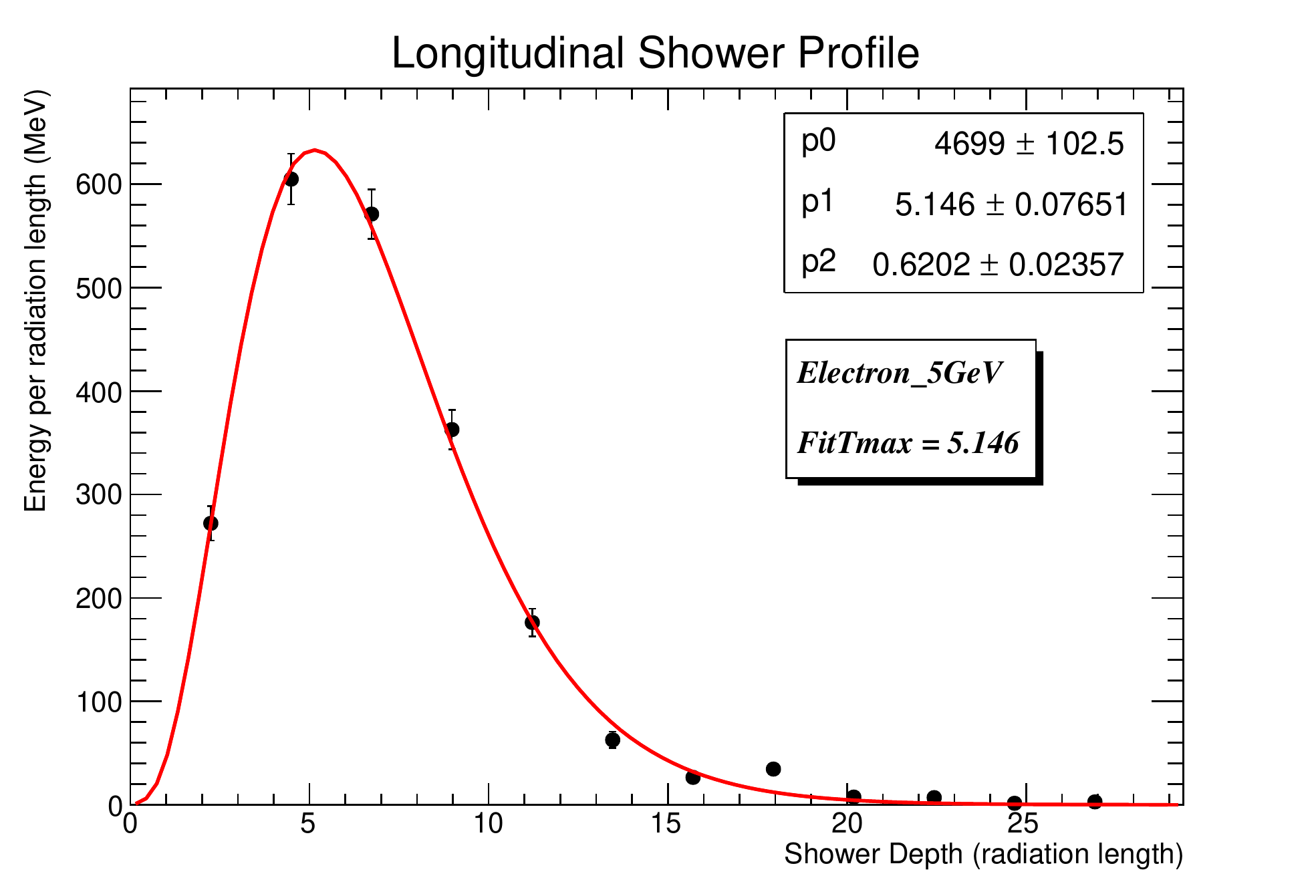}
  }
  \subfloat[50 GeV MC electron]{
    \includegraphics[width=.40\textwidth]{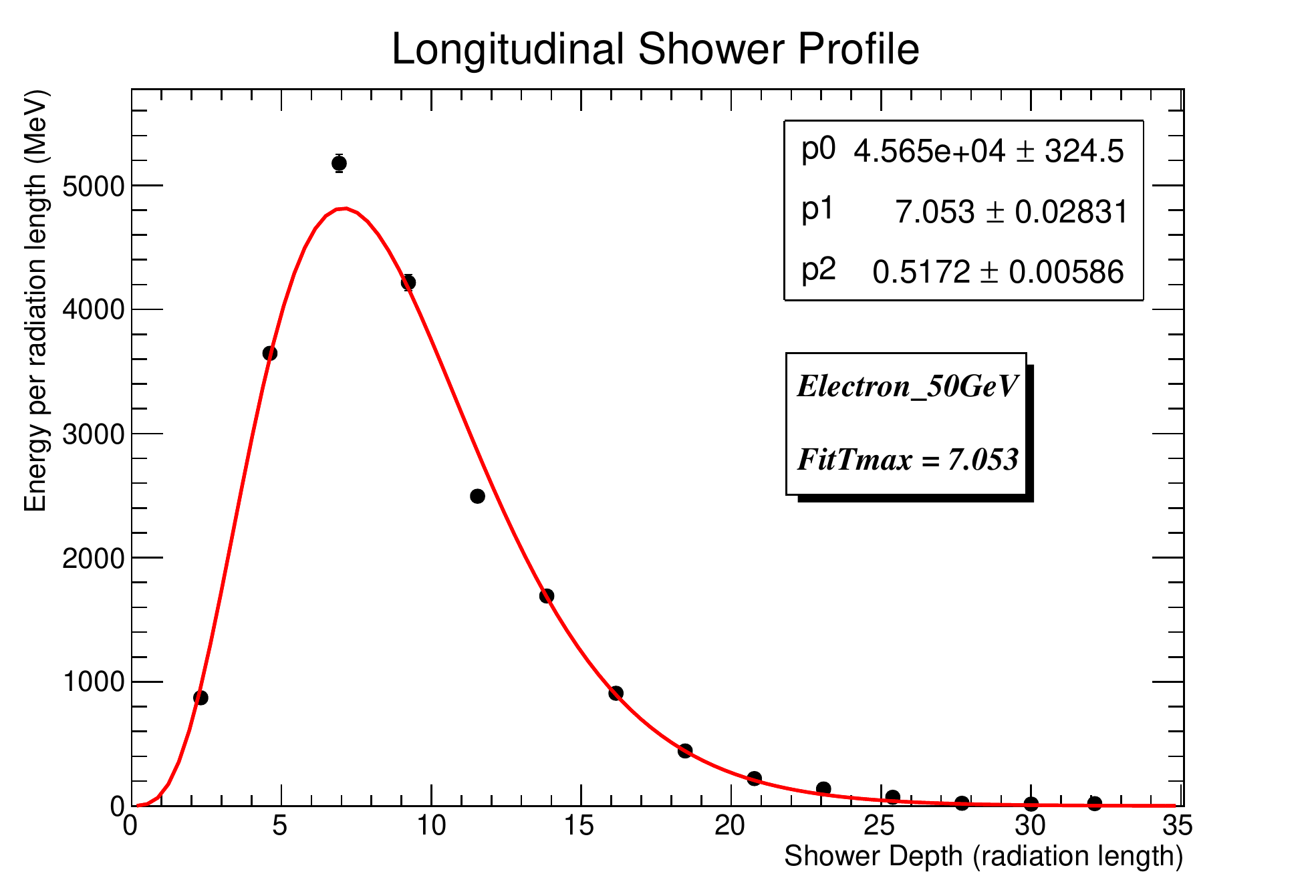}
  }\\
  \subfloat[500 GeV MC electron]{
    \includegraphics[width=.40\textwidth]{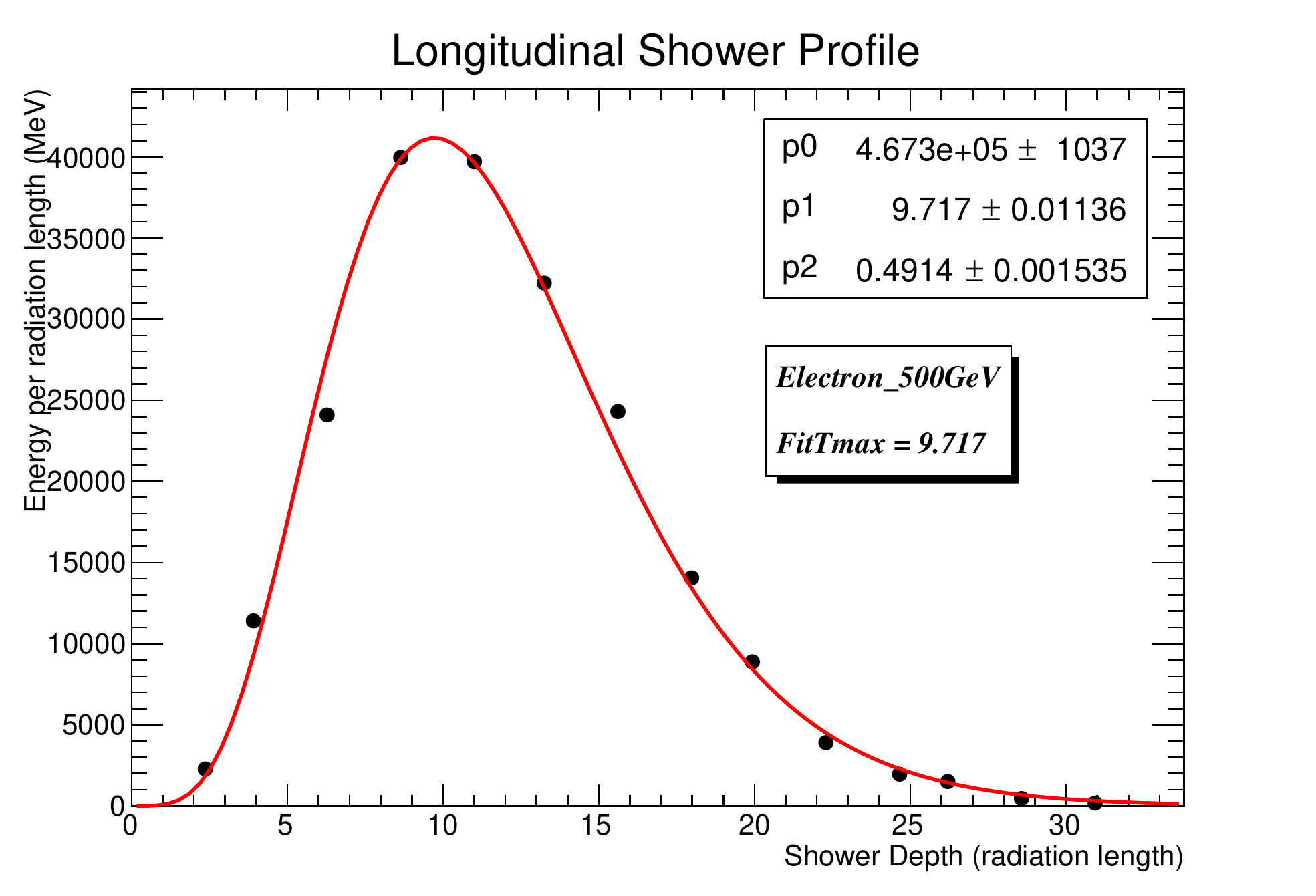}
  }
  \subfloat[5 TeV MC electron]{
    \includegraphics[width=.40\textwidth]{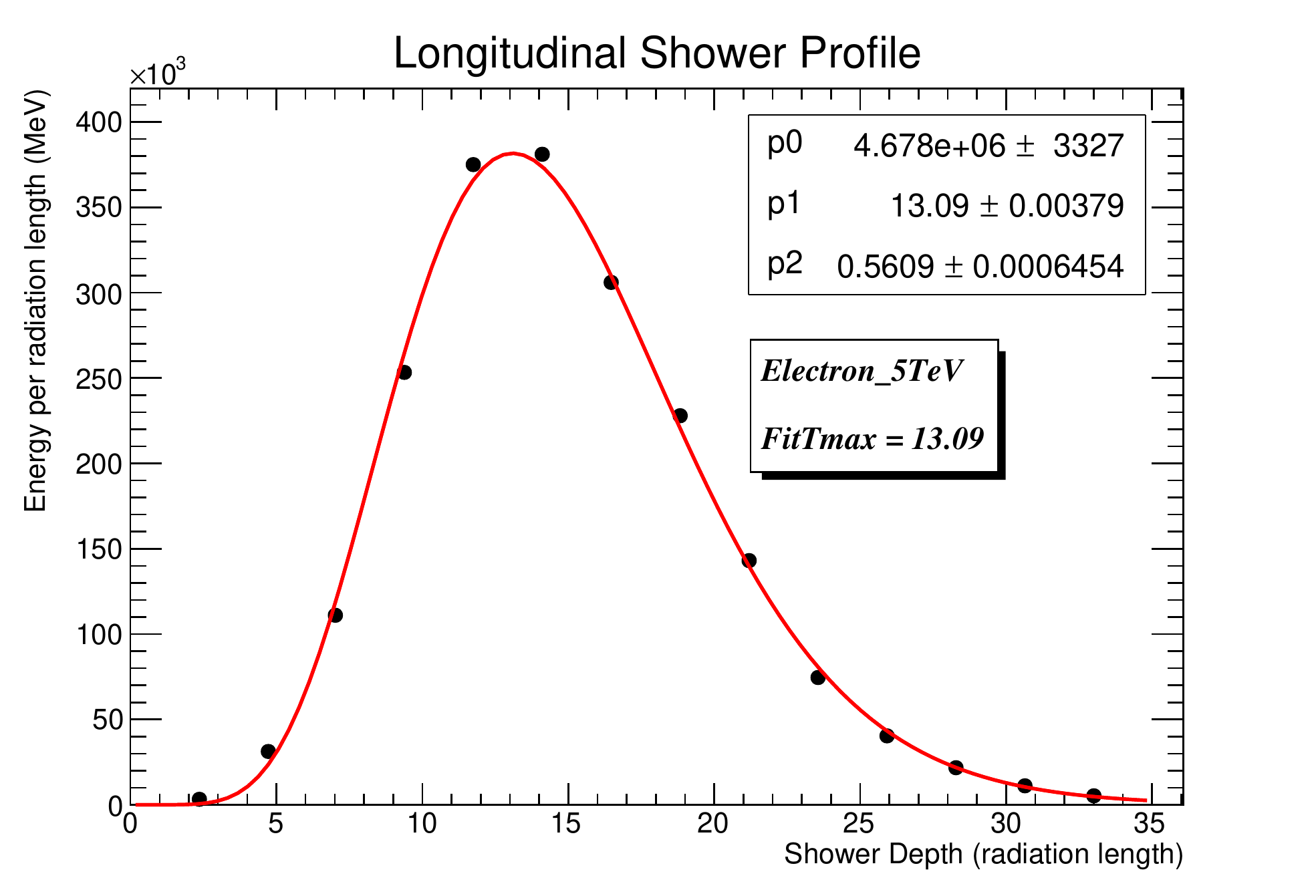}
  }
  \caption{Typical longitudinal shower profiles of (a)5GeV, (b)50GeV, (c)500GeV and (d)5TeV MC electrons. The profiles are fitted by the gamma-distribution function and the $T_{max}$ in units of radiation length is displayed on each canvas.}
  \label{fig-gammafit}
\end{figure*}

\begin{figure*}[!ht]
  \centering
  \subfloat[10 GeV MC electron]{
    \includegraphics[width=.32\textwidth]{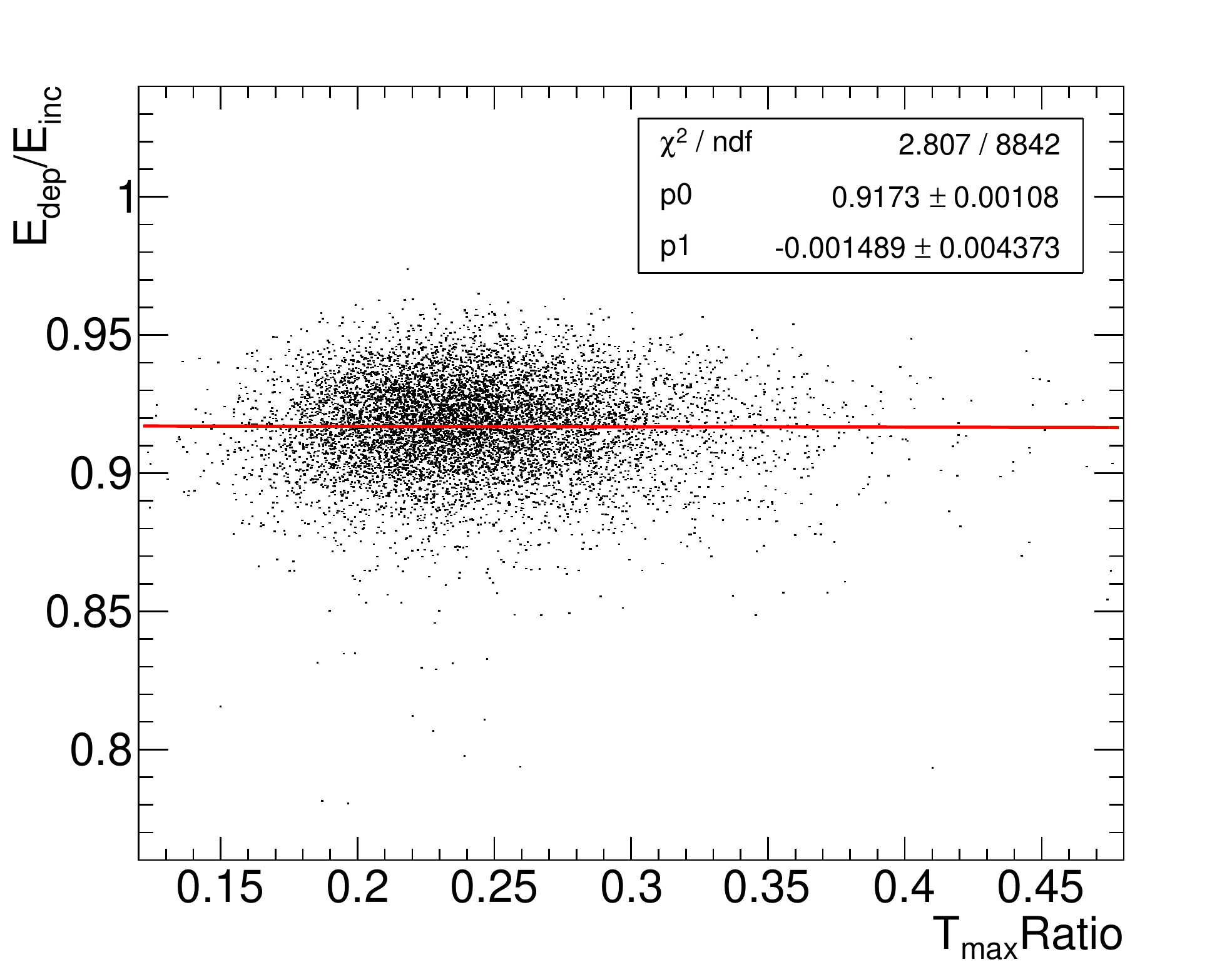}
  }
  \subfloat[100 GeV MC electron]{
    \includegraphics[width=.32\textwidth]{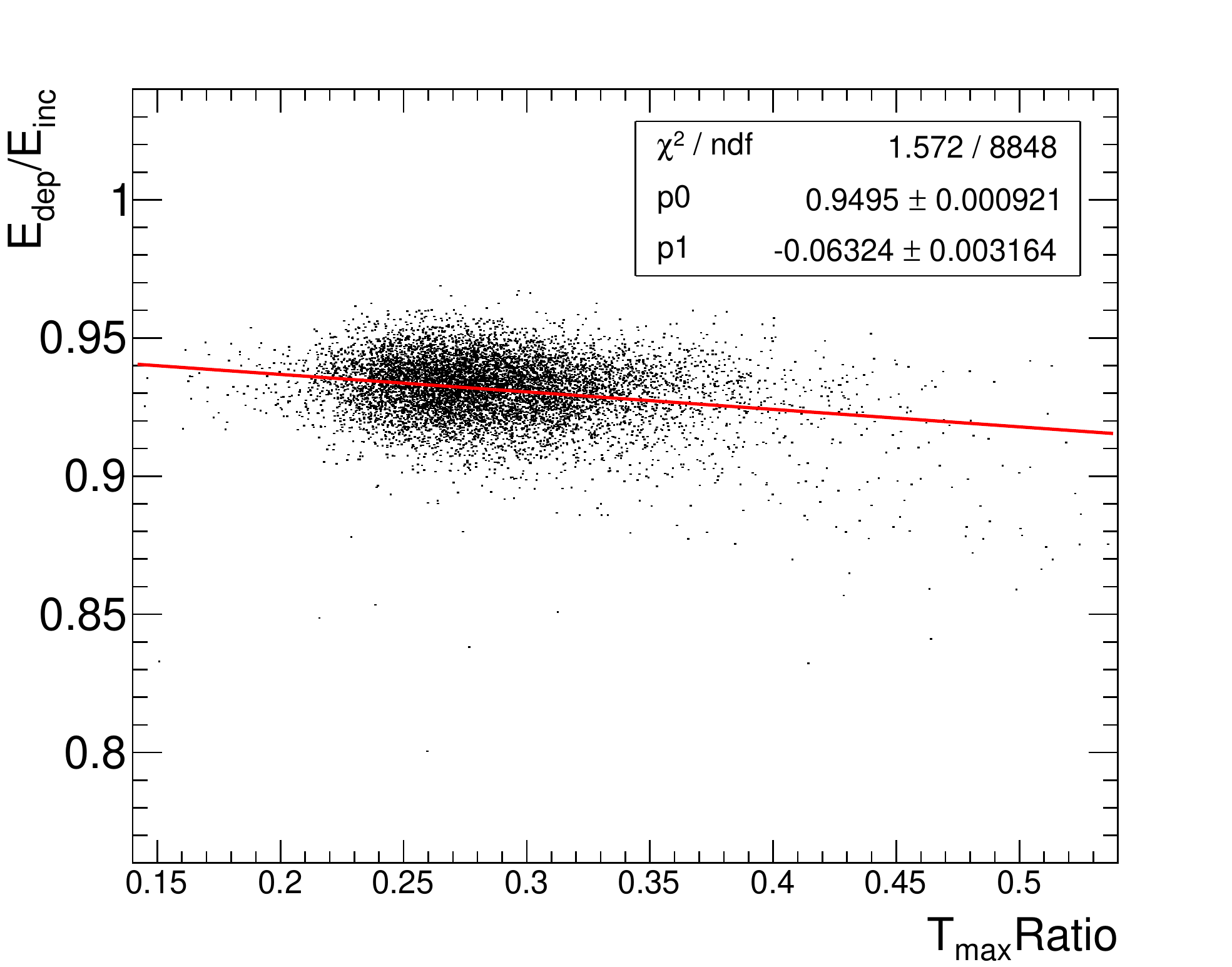}
  }
  \subfloat[1000 GeV MC electron]{
    \includegraphics[width=.32\textwidth]{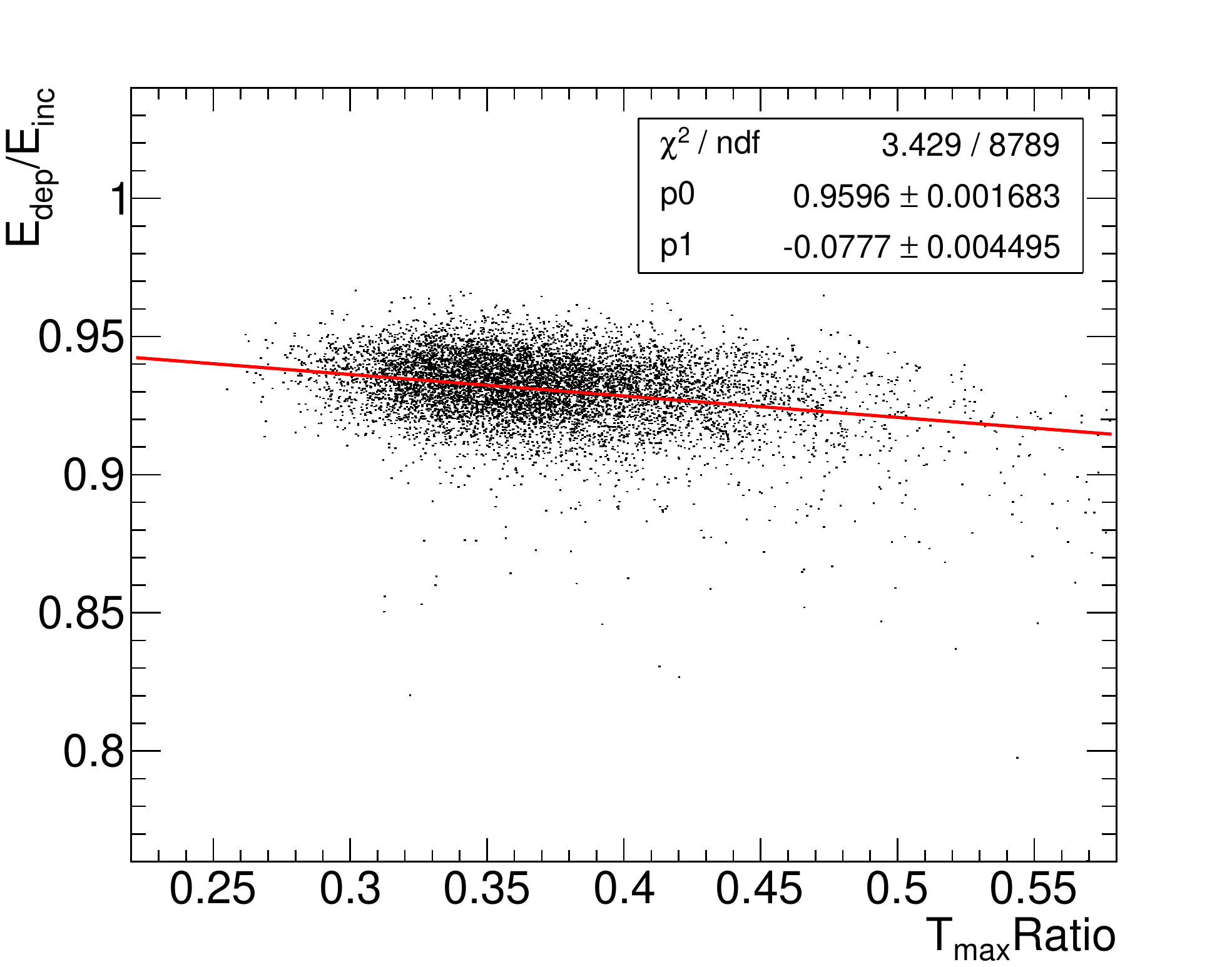}
  }
  \caption{The variations of $EdepRatio$ along with $T_{max}Ratio$ for (a)10GeV, (b)100GeV and 1000GeV MC electrons. The linearity relation between $EdepRatio$ and $T_{max}Ratio$ are fitted as red line.}
  \label{fig-tmratio}
\end{figure*}

The longitudinal segmentation of the ECAL allows a fit of the longitudinal shower profile, which provides a good way to correct the longitudinal energy leakage when the shower is not fully contained in the calorimeter. The longitudinal shower profile can be well described by a gamma-distribution function \cite{Longo1975} as the following: 
\begin{linenomath}
\begin{equation}
\frac{dE(t)}{dt} = E_{0} \cdot \frac{(\beta{t})^{\alpha-1} \cdot \beta \cdot e^{-\beta{t}}}{\Gamma(\alpha)}
\label{eq-gammadis}
\end{equation}
\end{linenomath}
where t is the shower depth in units of radiation length $X_{0}$, $\alpha$ is the shape parameter and $\beta$ is the scaling parameter. The depth of the shower maximum depends on $\alpha$ and $\beta$ according to:
\begin{linenomath}
\begin{equation}
T_{max} = \frac{\alpha-1}{\beta} \quad \quad T_{max}Ratio = \frac{T_{max}}{T_{total}}
\label{eq-tmax}
\end{equation}
\end{linenomath}
$T_{max}$ is the depth of the shower maximum and $T_{total}$ is the total length that the particle travels through. Note that here we just need to obtain a correct estimation of $T_{max}$ by fitting each individual profile with the gamma distribution. Fig.\ref{fig-gammafit} shows the fit results of typical longitudinal shower profiles for 5GeV, 50GeV, 500GeV and 5TeV MC electrons. It is clear that the particle with larger incident energy would have deeper shower maximum, which indicates that the rear energy leakage becomes more and more important with the increase of incident energy. For the electrons with a certain incident energy,  $T_{max}Ratio$ shows significant uncertainty form event to event (see Fig.\ref{fig-tmratio}) and it is found that $T_{max}Ratio$ has a good correlation with the energy deposition ratio. The bigger $T_{max}Ratio$ suggests the particle produces a posterior shower, which could cause more energy loss in gaps between the layers and indicate more rear energy leakage from backend of the calorimeter.

\subsection{Parameter Analysis}
\begin{figure*}[!ht]
  \centering  
  \includegraphics[width=1.0\textwidth]{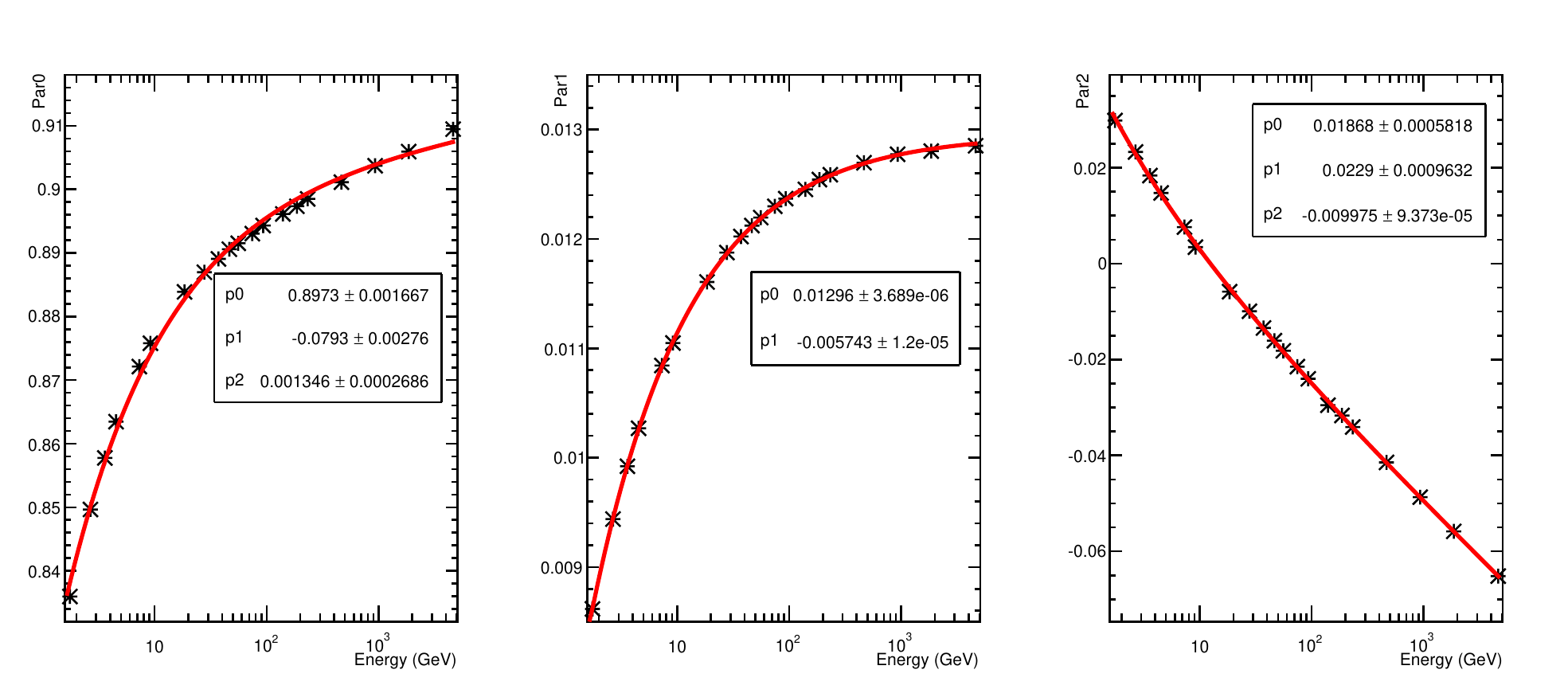}
  \caption{The parameterized results of three correction parameters. The three parameters are fitted by Equation.\ref{eq-parfit}}
  \label{fig-parameters}
\end{figure*}

Based on the lateral variable $CoreEneRatio$ and the longitudinal variable $T_{max}Ratio$ mentioned above, a bivariate function used for energy correction is proposed as the following:
\begin{linenomath}
\begin{equation}
\begin{split}
& EdepRatio  = \frac{E_{dep}}{E_{inc}} 
\\
& = p_{0} + p_{1} \cdot CoreEneRatio + p_{2} \cdot T_{max}Ratio
\end{split}
\label{eq-correction}
\end{equation}
\end{linenomath}
where $E_{dep}$ is the total deposited energy in the calorimeter and $E_{inc}$ is the incident energy. By fitting the $EdepRatio$ with the bivariate function for MC electrons with different incident energies, the correction parameters can be obtained. Fig.\ref{fig-parameters} shows how the parameters vary with the average deposited energy. They are parameterized by different energy-dependent empirical functions as the following:
\begin{linenomath}
\begin{equation}
\begin{split}
p_{0}(E) & = k_{0} + \frac{k_{1}}{\sqrt{E}} + k_{2}ln(E)
\\
p_{1}(E) & = k_{0} + \frac{k_{1}}{\sqrt{E}}
\\
p_{2}(E) & = k_{0} + \frac{k_{1}}{\sqrt{E}} + k_{2}ln(E)
\end{split}
\label{eq-parfit}
\end{equation}
\end{linenomath}
The energy correction process for each electromagnetic shower can be described as the following: firstly calculating $CoreEneRatio$ with the energy deposition in each crystal and the measured incident angle, then fitting the longitudinal shower profile to obtain $T_{max}Ratio$, and finally estimating the incident energy with Equation.\ref{eq-correction}, $E_{cor}=E_{dep}/EdepRatio$.

\begin{figure*}[!ht]
  \centering  
  \includegraphics[width=.80\textwidth]{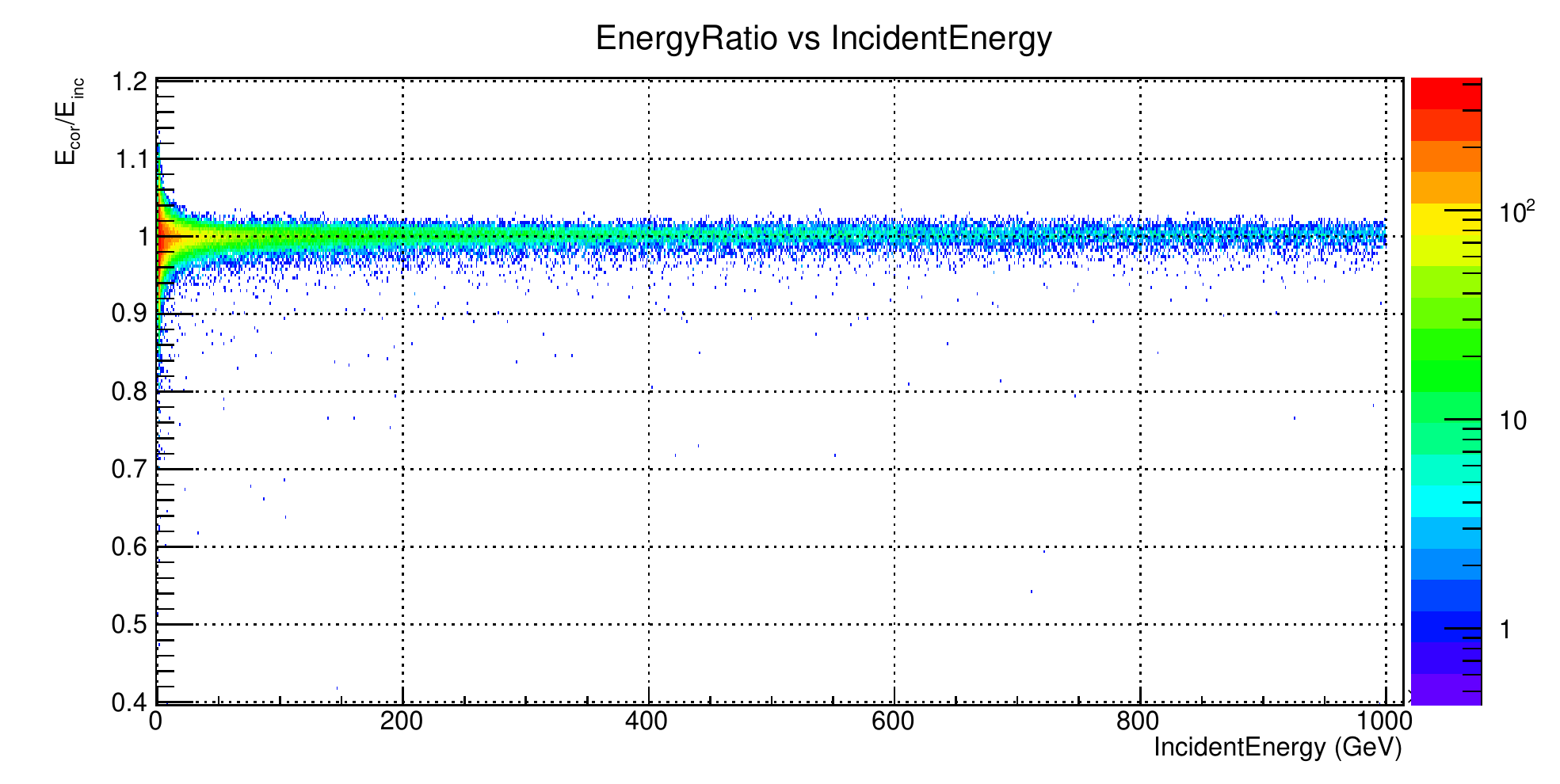}
  \caption{The relation between EcorRatio and the incident energies, which indicates that the correction method has weak dependence with particle's energy. }
  \label{fig-enegrydependence} 
\end{figure*}

\begin{figure*}[!ht]
  \centering  
  \includegraphics[width=.80\textwidth]{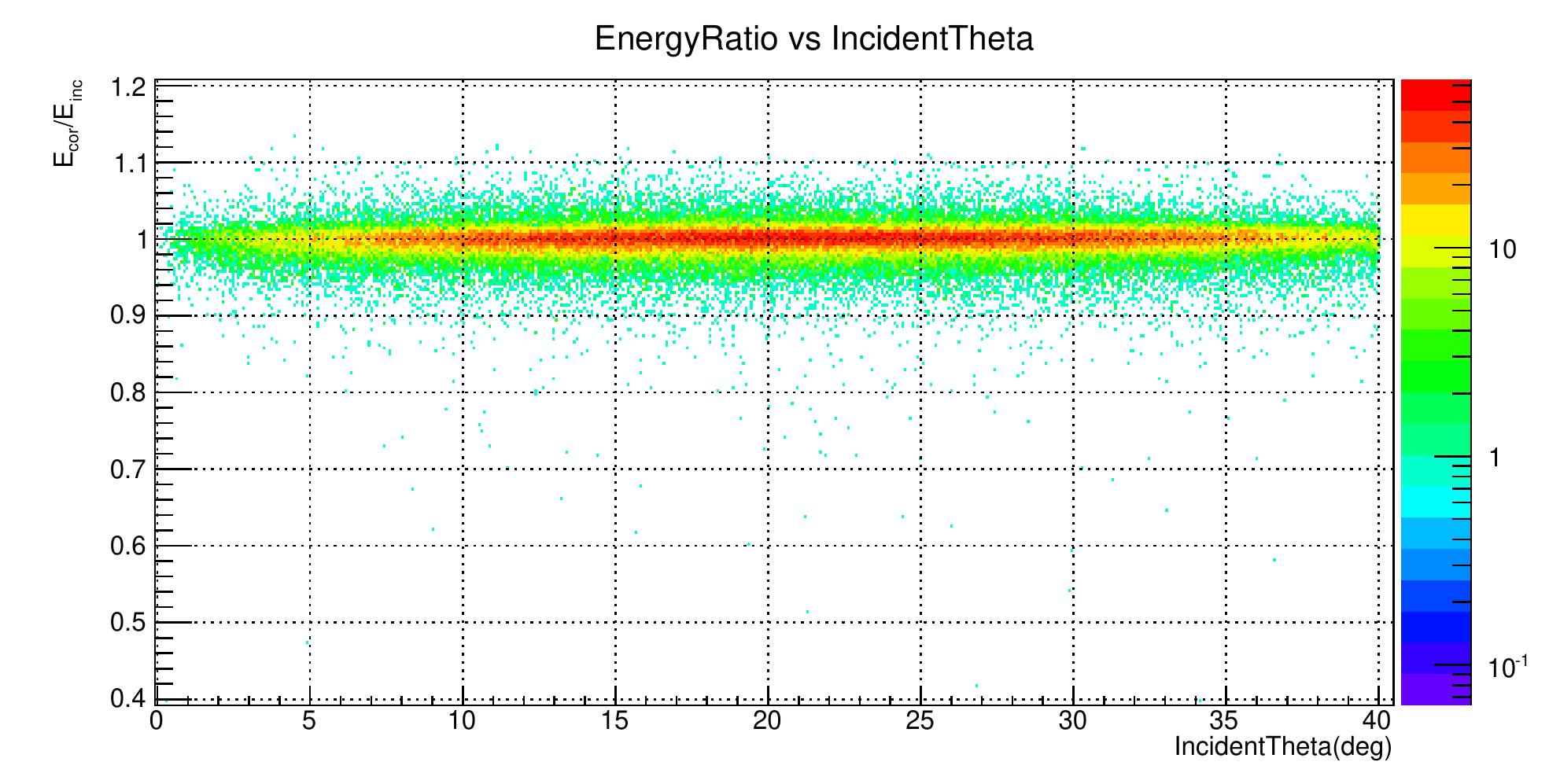}
  \caption{The relation between EcorRatio and the incident angles, which indicates that the correction method has weak dependence with particle's direction. Theta means the angle between the incident track and the z-axis.}
  \label{fig-angledependence}
\end{figure*}

To check this correction method's dependence with the energy and direction of incident particles, an isotropic electron sample with a power-law spectrum of $E^{-1}$ has been generated with the DAMPE simulation package. Fig.\ref{fig-enegrydependence} and Fig.\ref{fig-angledependence} show the variation of $EcorRatio$ ($E_{cor}/E_{inc}$) with incident energy and incident angle respectively. The $EcorRatio$ remains around 1.0 with the variation of energy or angle, indicating that our method is independent of these two quantities.

\section{Performance studies for Electron Beam Test}
\begin{figure*}[!ht]
  \centering
   \includegraphics[width=1.0\textwidth]{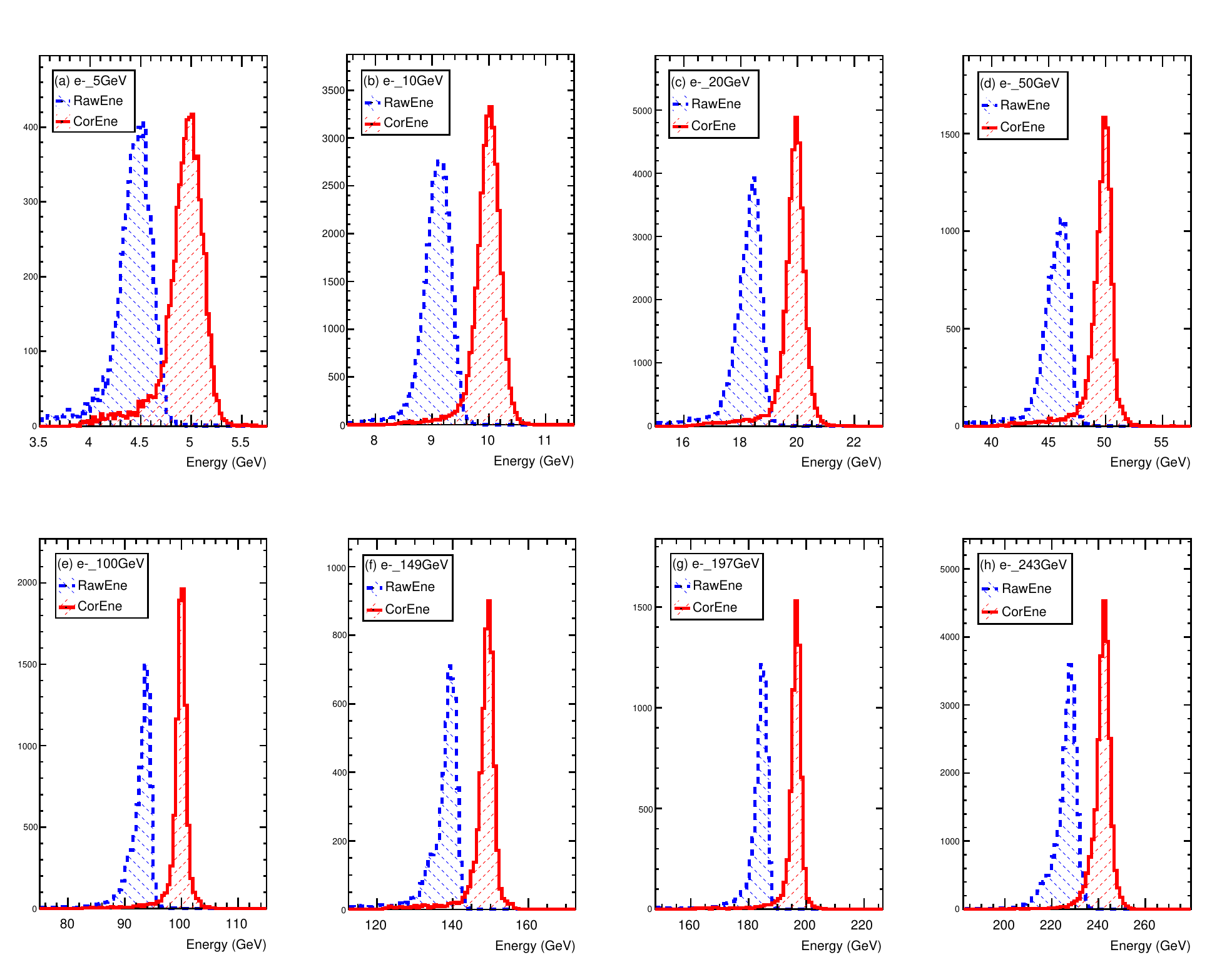}
  \caption{The measured energy distributions before (blue) and after (red) correction for the normal-incident electron beams with certain energies from 5 GeV to 243 GeV.}
  \label{fig-beamtest}
\end{figure*}

The electron beam test of DAMPE was carried out at CERN-PS T9 and CERN-SPS H4 from October to November of 2014 for the energy range from 0.5 GeV to 243 GeV \cite{ZhangZY2016}. The beam test data were reprocessed applying the latest reconstruction and selection algorithms which are same as used in flight-data analysis. Fig.\ref{fig-beamtest} shows the results of the correction method applied to the normal-incident electron beams of different incident energies (1 $\sim$ 243 GeV). It is evident that the energy resolutions for electron beams above 5 GeV are all significant improved after correction. Meanwhile, the bias of the deposited energy distribution has been adequately corrected as well, thereby optimizing the energy measurement.

\begin{figure*}[!ht]
  \centering
   \includegraphics[width=0.8\textwidth]{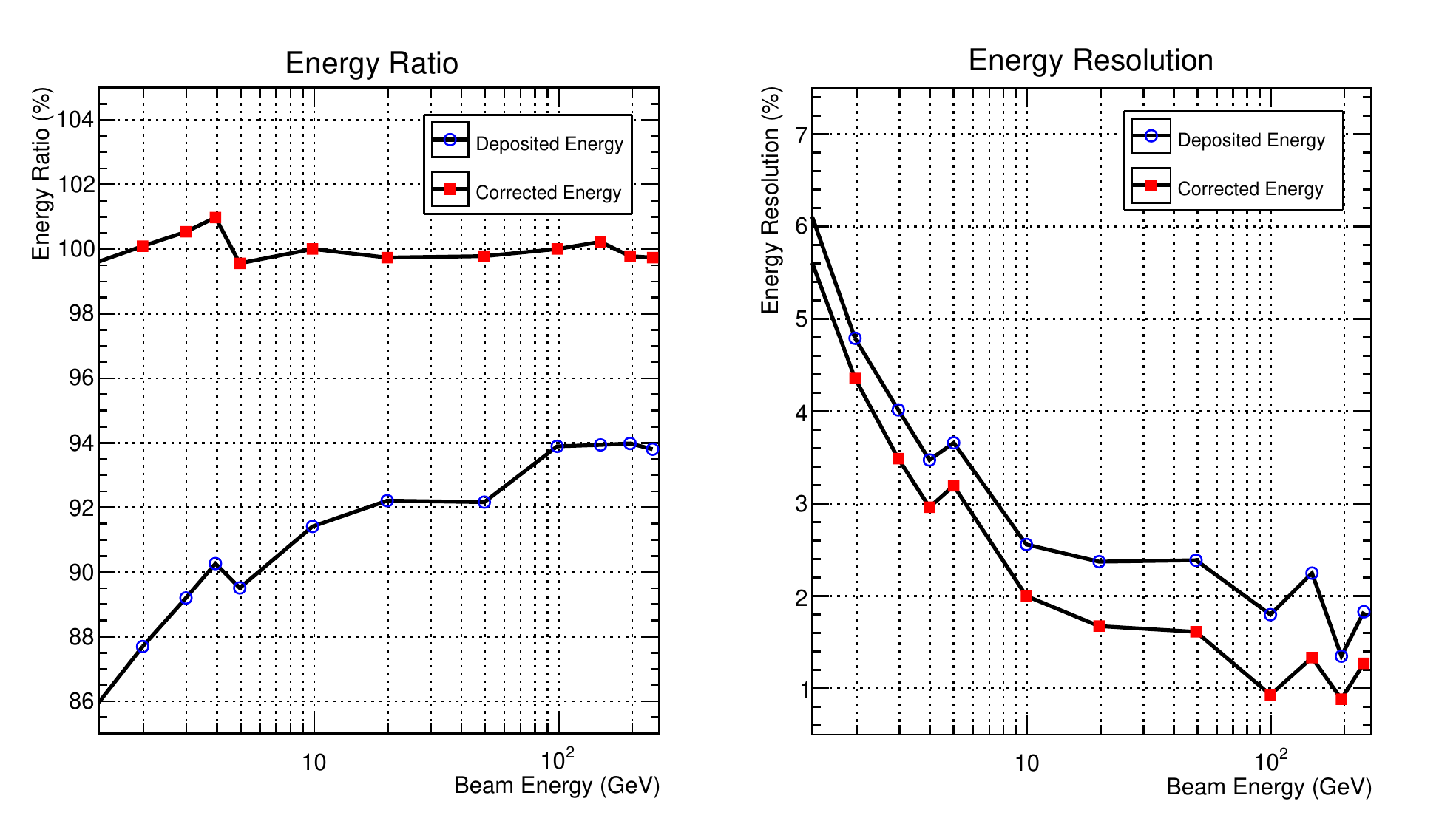}
  \caption{Energy Ratio (left) and Energy Resolution (right) results for electron beams before (blue) and after (red) correction.}
  \label{fig-finalresult}
\end{figure*}

Fig.\ref{fig-finalresult} shows the energy ratios (left) and the energy resolutions (right) before and after correction.The energy ratio is defined as $E_{dep}/E_{beam}$ or $E_{cor}/E_{beam}$ ($E_{dep}$ or $E_{cor}$ means the peak value of the energy distribution). The $E_{cor}/E_{beam}$ shows slight fluctuation around $100\%$ and the linearity is kept within $1\%$ in the energy range from 1 to 243 GeV. The energy resolution is defined from the energy distribution as the half width of the smallest window that contains the 68\% of the events \cite{Fermi2011}. The energy resolutions after correction, varying between 5.38\% at 1GeV and 0.87\% at 197GeV, show evident improvement compared with raw deposited energy. 

\section{Conclusions}
Based on the detailed MC simulation package of DAMPE (see section.2), a parameterized energy correction method for the electromagnetic showers produced by $e^{\pm}/{\gamma}$ in BGO-ECAL has been studied. The lateral and longitudinal information of the shower development is used to construct the correction function, simultaneously taking into account energy losses due to dead material and longitudinal energy leakage. The correction parameters are parameterized by different energy-dependent empirical functions in the energy range from 1 GeV to 5 TeV. The energy correction method shows no dependence on the primary energy or direction of incident particle. In principle, this correction method could be not only applied to DAMPE, but extended to any other experiment equipped with a electromagnetic calorimeter. The results of this correction method applied to data from an electron beam test at CERN show encouraging improvements of the energy linearity and resolution. The correction has been applied in flight-data analysis for electron cosmic-rays and $\gamma$-rays.

\section{\it Acknowledgments.}
The authors would like to thanks Chinese and European colleagues of DAMPE collaboration for useful discussions. The authors acknowledge the generosity of CERN for providing beam time allocation and technical assistance at the PS and SPS beam lines, as well as general logistical support. This work was supported by the National Key Research and Development Plan under Grant No. 2016YFA0400200, the Strategic Priority Research Program on Space Science of the Chinese Academy of Science under Grant No. XDA04040400, the National Natural Science Foundation of China under Grants No. 11303105, No. 11303107 and No. 11673075. 

\bibliography{mybib}

\end{document}